\documentclass[prb,aps,twocolumn,showpacs,eqsecnum,superscriptaddress,twoside]{revtex4-1}
\usepackage{amsmath}
\usepackage{amssymb}
\usepackage{stmaryrd}
\usepackage{graphicx,color}
\begin{document}

\title{Microwave realization of quasi-one-dimensional systems with correlated
disorder}

\author{O.~Dietz}
\affiliation{Fachbereich Physik, Philipps-Universit\"{a}t
Marburg, Renthof 5, D-35032 Marburg, Germany}
\email{otto.dietz@physik.uni-marburg.de}

\author{U.~Kuhl}
\affiliation{Fachbereich Physik, Philipps-Universit\"{a}t
Marburg, Renthof 5, D-35032 Marburg, Germany}
\affiliation{Laboratoire de Physique
de la Mati\`{e}re Condens\'{e}e, CNRS UMR 6622, Universit\'{e} de Nice
Sophia-Antipolis, F-06108 Nice, France} \email{ulrich.kuhl@unice.fr}

\author{H.-J.~St\"{o}ckmann}
\affiliation{Fachbereich Physik,
Philipps-Universit\"{a}t Marburg, Renthof 5, D-35032 Marburg, Germany}

\author{N.~M.~Makarov}
\affiliation{Instituto de Ciencias, Universidad
Aut\'{o}noma de Puebla, Priv. 17 Norte No 3417, Col. San Miguel Hueyotlipan,
Puebla, Puebla, 72050, Mexico}

\author{F.~M.~Izrailev}
\affiliation{Instituto de F\'{\i}sica, Universidad
Aut\'{o}noma de Puebla, Apartado Postal J-48, Puebla, Puebla, 72570, Mexico}

\date{\today}

\begin{abstract}
A microwave setup for mode-resolved transport measurement in quasi-one-dimensional (quasi-1D)
structures is presented. We will demonstrate a technique for direct measurement of
the Green's function of the system. With its help we will investigate quasi-1D structures with various types of disorder. We will focus on stratified structures, i.e., structures that are homogeneous perpendicular to the direction of wave propagation. In this case the interaction between different channels is absent, so wave propagation occurs individually in each open channel. We will apply analytical results developed in the theory of one-dimensional (1D) disordered models in order to explain main features of the quasi-1D transport. The main focus will be selective transport due to long-range correlations in the disorder. In our setup, we can intentionally introduce correlations by changing the positions of periodically
spaced brass bars of finite thickness. Because of the equivalence of the stationary Schr\"{o}dinger equation and the Helmholtz equation, the result can be directly applied to selective electron transport in nanowires, nanostripes, and superlattices.
\end{abstract}

\pacs{72.15.Rn, 42.25.Bs, 42.70.Qs}

\maketitle

\section{Introduction}
Because of progress in nano- and materials science, the investigation of (electron) wave
propagation in periodic one-dimensional (1D) systems has recently attracted
growing attention (see, e.g., Ref.~\onlinecite{mar08} and references therein).
Systems of particular interest are metamaterials or stacked layer structures, as
well as superlattices.
One of the important problems that is still far from being completely understood
is the role of disorder due to fluctuations in the width of layers or due to
variations in the material parameters, such as the dielectric constant, magnetic
permeability, or barrier height (for electrons)
\cite{bal85,mcg93,smi00,she01,par03,esm06,don06,nau07,pon07,asa07,izr10a,asa10,mog10}.
This disorder was seen mainly as an unwanted side effect in experimental
realization \cite{izr10a}. But with enhanced ability to control these features,
disorder itself becomes an interesting candidate for targeted manipulation of
transport properties.
In particular, the correlations in the disorder may exhibit unusual features. One
of the earliest findings was that short range correlations can inhibit
localization in low dimensional systems \cite{dun90,bel99}.

Later, it was shown, both theoretically
\cite{izr99,kro99,izr01a,her08,izr01b,izr03d,izr05a,die08d,izr10b} and experimentally
\cite{kuh00a,kuh08a,die10d,kuh10b}, that specific long-range correlations can even be used to
significantly enhance or suppress the wave localization in any desired windows of
frequency. Recent analytical studies extended the theoretical predictions from
purely 1D models to quasi-one-dimensional (quasi-1D) randomly stratified structures
\cite{izr04a,izr05a}, as well as to random, periodic-on-average stacked systems
with bilayer unit cells \cite{mak07c,izr09,lun09a,lun09b}. These studies provide
the application of correlated disorder to photonic crystals, metamaterials and
layered structures, e.\,g., semiconductor superlattices.

In this paper we explore the validity and applicability of the theoretical results
by studying the microwave transmission through a quasi-1D waveguide open on both ends.
It was shown in Ref.~\onlinecite{tud08} that
the transport properties can be effectively studied by measurement of the
Green's function, up to a factor describing the antenna properties. This method
can be considered as an essential tool to analyze the experimental data. It allows
us to test experimentally the analytical predictions from Ref.~\onlinecite{tud08}
in open waveguides with and without disorder. Furthermore, we extend the
experimental confirmation of selective transport to quasi-1D systems that emerge
because of long-range correlations. Many studies on 1D systems with correlated
disorder involve hundreds of scatterers \cite{bel99,kuh00a,kuh08a}. In present
experiments only 26 scatterers are used. We demonstrate that this small number is
already sufficient to reveal the predicted effect of correlated disorder.

\section{Experimental Setup and Basic Equations}

\begin{figure}
\includegraphics[width=\columnwidth]{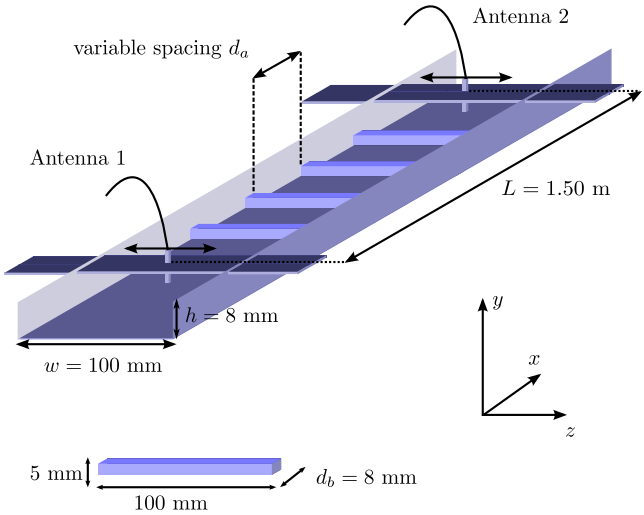}
\caption{\label{fig:sketch} (Color online)
Not-to-scale sketch of experimental setup with top plate removed. Brass-bar inlays can
be freely placed inside the waveguide. Number of bars varies according to spacing $d_a$.
Antennas shown here with parts of the top plate can be moved
separately along the $z$ direction. Distance along the $x$ axis between antennas
is $L=1.50$\,m; total waveguide length is 2.38\,m.
Absorbers at the ends of the waveguide are omitted.}
\end{figure}

In our study we use an experimental setup that allows us to measure microwave
transport through a multimode metallic-wall waveguide with an internal structure
of different kinds (see Fig.~\ref{fig:sketch}). The input
(transmitting) and output (receiving) antennas are plugged into the waveguide via
slides that can be shifted stepwise by two motors. The absorbers are placed on
both ends of the waveguide, in order to reduce the reflection from the ends, i.e.,
to consider the setup as a \emph{finite waveguide with open ends}. The values of
the height $h$, width $w$ and the distance between the antennas -- the effective
length $L$ of the waveguide -- meet the conditions
\begin{equation}\label{WaveguideDef}
  h\ll w\ll L
\end{equation}
that are important for a proper theoretical analysis.

In an empty rectangular waveguide the electromagnetic field can be separated into
two components of different polarization: the TM polarization with transverse
magnetic field and TE polarization with transverse electric field \cite{jac62}. In
our experiments all measurements are performed below the cut-off frequency
$\nu_\textrm{cut}=c/2h\approx18.75$\,GHz. In this case only the lowest
TE-component, having the electric field stretched along the $y$ axis
and the magnetic field being in the $(xz)$ plane (see the coordinate system in
Fig.~\ref{fig:sketch}), can propagate. Therefore, our experimental setup within the frequency
interval $\nu<\nu_\textrm{cut}$ and with the conditions in Eq.~\eqref{WaveguideDef} can
be considered as a quasi-1D open-strip structure, occupying the region
\begin{equation}\label{Q1Ddef}
  0\leqslant x\leqslant L,\qquad\qquad 0\leqslant z\leqslant w.
\end{equation}

In order to develop a proper theory for our model, we treat the input antenna 1 in
$(x,z)$ plane as a point source, while the output antenna 2 is an observation
point. As is known, the electric field of a point source is determined by the
retarded Green's function ${\cal G}(|x-x'|;z,z')$. In Ref.~\onlinecite{tud08} it
was shown that for the correct analysis of transport one has to take into account
the influence of coupling between the antennas and the waveguide. As a result, the
Green's function for the empty waveguide in the normal-mode representation reads
\begin{eqnarray}\label{Gmode}
{\cal G}(|x-x'|;z,z')&=&
\sum_{n=1}^{N_w}A_n\sin\left(\frac{\pi nz}{w}\right) \sin\left(\frac{\pi nz'}{w}\right)\nonumber\\[6pt]
&&\times\frac{\exp\left(ik_n|x-x'|\right)}{ik_nw}.
\end{eqnarray}
Here we introduced the factor $A_n$ which effectively describes the
coupling between the antennas and normal modes labeled by index $n$. In our
experiment this factor $A_n$ is regarded as a fitting parameter. Without antenna
coupling ($A_n=1$) Eq.~\ref{Gmode} coincides with the Green's function of the
two-dimensional Helmholtz equation complemented by the Dirichlet boundary conditions at
metallic walls of the waveguide.

The quantities $\pi n/w$ and $k_n$ are, respectively, the discrete values of
transverse, $k_z$, and longitudinal, $k_x$, wave numbers:
\begin{eqnarray}\label{kn}
k_n&=&\sqrt{k^2-(\pi n/w)^2}= \frac{2\pi}{c}\sqrt{\nu^2-(n\nu_c)^2},\nonumber\\[6pt]
&&n=1,2,3,\ldots,N_w.
\end{eqnarray}
Here $k=2\pi\nu/c$ is the total wave number for the electromagnetic
wave of frequency $\nu$. For the modeling of electron quasi-1D structures (such as
nanowires and nanostripes) the wave number $k$ should be regarded as the Fermi wave
number within the isotropic Fermi-liquid model.

The total number $N_w$ of propagating waveguide modes (for which $k_n$ is real) is
determined by the integer part $\llbracket\ldots\rrbracket$ of the mode parameter
$kw/\pi=\nu/\nu_c$,
\begin{equation}\label{Nw}
  N_w=\llbracket kw/\pi\rrbracket=\llbracket\nu/\nu_c\rrbracket.
\end{equation}
Correspondingly, the sum in Eq.~\eqref{Gmode} runs only over the propagating modes with $n\leq N_w$
and ignores the contribution of evanescent modes for which $n>N_w$. Indeed, for
evanescent modes the values of $k_n$ are purely imaginary; therefore, they do not
contribute to the transport. The critical frequency $\nu_c$,
\begin{equation}\label{eq:cutoff}
  \nu_c=c/2w,
\end{equation}
below which there are
only evanescent modes ($kw/\pi<1$, i.e., $N_w=0$) is called the cut-off
frequency. Evidently, there is no transport for $\nu<\nu_c$. Note that here
$\nu_c$ is the cut-off frequency for the first mode with $n=1$. The cut-off
frequency $\nu_c^{(n)}$ for each higher mode with index $n$ is defined as
$\nu_c^{(n)}=n\nu_c$.

\section{Scattering Matrix}

It is suitable to define the scattering matrix $S_{nm}$ for the whole
waveguide ($|x-x'|=L$) as the twofold sine-Fourier transform,
\begin{equation}\label{eq:Snm-def}
S_{nm}=\frac{2}{w}\int_0^wdzdz'
\sin\left(\frac{\pi nz}{w}\right) \sin\left(\frac{\pi mz'}{w}\right) {\cal
G}(L;z,z').
\end{equation}
One can see that in accordance with Eq.~\eqref{Gmode}
the scattering matrix for the empty quasi-1D waveguide [Eq.~\eqref{Q1Ddef}] is diagonal
in the mode representation. In this case the $S$ matrix is proportional to the
Green's function for free 1D propagation with the wave number $k_n$:
\begin{equation}\label{eq:Snm-dg}
S_{nm}=A_n\frac{\exp\left(ik_nL\right)}{2ik_n}\Theta(N_w-n)\delta_{nm}.
\end{equation}
Here $\Theta(x)$ stands for the Heaviside unit-step function, where
$\Theta(x<0)=0$ and $\Theta(x\geqslant0)=1$. The result [Eq.~\eqref{eq:Snm-dg}] is quite
natural, since in an empty waveguide there are no transitions between the normal
modes. Therefore, the transport through any channel can be considered
independently of other channels.

Below we call the elements $S_{nm}$ of the scattering matrix [Eq.~\eqref{eq:Snm-def}]
the scattering amplitudes. Then, the scattering probability $T_{nm}$
from the mode $n$ to mode $m$, the mode-transport coefficient $T_n$ (for the
scattering of a given $n$th ``incoming'' mode into all ``outgoing'' modes) and the
total transport coefficient $T$ are naturally defined as follows:
\begin{equation}\label{eq:Tnm-def} T_{nm}=|S_{nm}|^2,\quad
T_n=\sum_{m=1}^{N_w}T_{nm},\quad T=\sum_{n=1}^{N_w}T_{n}.  \end{equation}

\begin{figure}
\includegraphics[width=\columnwidth]{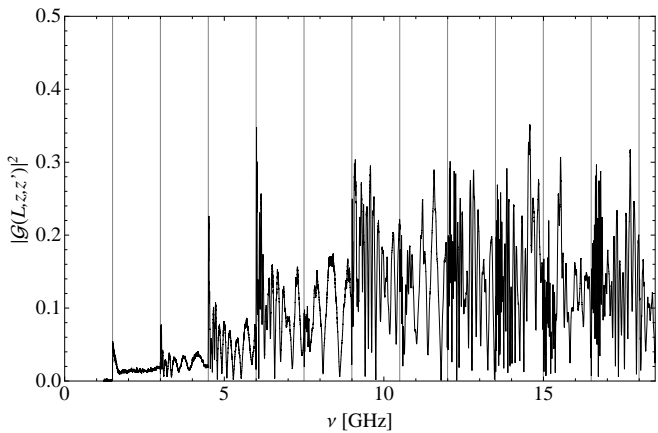}
\caption{\label{fig:EmptyWaveguide} (Color online)
Transmission spectrum $|{\cal G}(L;z,z')|^2$
through the empty waveguide for fixed antenna position $z$=$z'$=55\,mm. Vertical
lines show the cut-off frequencies $\nu_c^{(n)}$ for different modes.}
\end{figure}

One should stress that although the above definitions of $T_{nm}, T_n$ and $T$
correspond to the standard ones known in the theory of transport through quasi-1D
disordered structures (e.\,g.~see Ref.~\onlinecite{bee97}), here the physical meaning
of all these quantities is different. Indeed, the expression for the scattering
matrix $S_{nm}$ is originated from the single-particle Green's function, not from
the Kubo formula, which is based on the two-particle Green's function. The price
for this simplification can be directly seen from Eq.~(\ref{eq:Snm-dg}), which has a
divergence at values of $k_n=0$. Therefore, the transmission coefficients
[Eq.~(\ref{eq:Tnm-def})] can not be normalized. On the other hand, there
is no divergence in the standard expressions for scattering matrices. However, our
point is that this somewhat new definition of the scattering matrix, based on the
mode representation of the Green's function, has an advantage when we consider
setups similar to the one in our experiment. In spite of the absence of rigorous expressions
for the localization length developed in connection with this type of scattering
matrix, in the following we show that the knowledge of theoretical results
obtained for 1D disordered models very much helps us to understand the
transport properties of the experimental setup.

For the experimental data analysis, the continuous Fourier transform
[Eq.~\eqref{eq:Snm-def}] is replaced by its discrete counterpart, allowing us to compute
$S_{nm}$ from ${\cal G}(L;z,z')$ measured for different positions $z$ and $z'$ of
input and output antennas. A typical measurement series with the starting points
$z=z'=5$\,mm and step size $10$\,mm yields $10^2$ individual measurements. An
example of the transmission spectrum obtained in one measurement is shown in
Fig.~\ref{fig:EmptyWaveguide}.
It should be kept in mind that in our experiment the total transport
coefficient $T$ is always less than 1, since the antennas are not perfectly coupled
to the waveguide. Furthermore, at both ends about half the energy escapes toward
the open ends. The amount of energy lost is not accessible in the present setup.
Moreover, the absorption inside the waveguide reduces the value of $T$ further. Practically, in
our experiment the total transport coefficient $T$ never exceeds the value
$T=0.17$.

\begin{figure}
\includegraphics[width=\columnwidth]{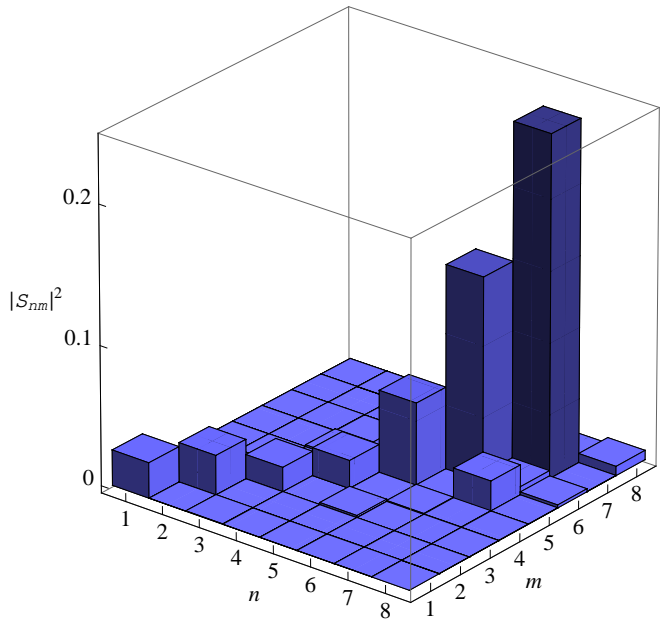}
\caption{\label{fig:emptywaveguide_snm} (Color online)
Modulus square of the scattering matrix elements
$T_{nm}=|S_{nm}|^2$ at $\nu=11$\,GHz with seven propagating modes.}
\end{figure}

The Fourier transform [Eq.~\eqref{eq:Snm-def}] can be performed for every single
frequency $\nu$ in the spectra yielding the scattering matrix $S_{nm}$ as a
function of $\nu$. An example of the scattering probability $T_{nm}=|S_{nm}|^2$
for an empty waveguide and fixed value $\nu=11$\,GHz is shown in
Fig.~\ref{fig:emptywaveguide_snm}. At this frequency there are seven propagating
modes. Since there are no scatterers in the waveguide, there should be no
transitions between different propagating modes, therefore, the matrix $T_{nm}$ is
expected to be diagonal in accordance with Eq.~\eqref{eq:Snm-dg}. Experimentally,
the total contribution of the off-diagonal elements in comparison with that of
diagonal elements is found to be always less than 5\,\% (typically much smaller).
The main source of these off-diagonal terms is the perturbation caused by
antennas. We always observe that the highest propagating mode gives the main
contribution to the transport, as in Fig.~\ref{fig:emptywaveguide_snm}, where
the term $|S_{77}|^2$ is the largest one. This fact is in complete correspondence
with the expectation from the theoretical predictions in the next section.

\section{Empty Waveguide}

\begin{figure}
\includegraphics[width=\columnwidth]{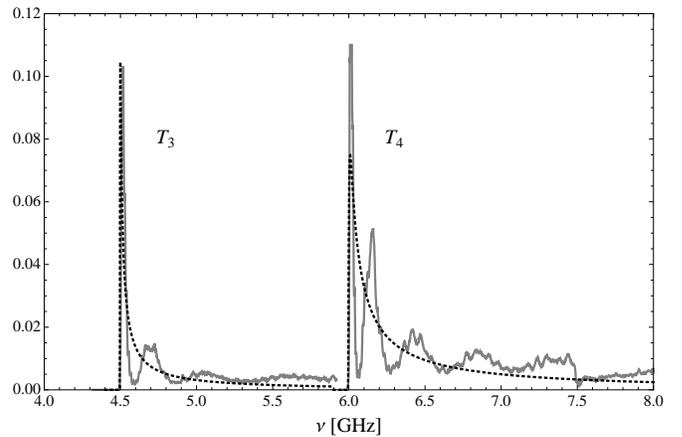}
\caption{\label{fig:emptywaveguide_mode} (Color online)
Mode propagation for the third and fourth modes
through the empty waveguide (solid line) compared to the theoretical prediction of
Eq.~\eqref{eq:Tn-EmW} (dotted line); here $|A_3|^2=30$ and Im$\,k=0.03$ for the
third mode and $|A_4|^2=146$ and Im$\,k=0.06$ for the fourth one.}
\end{figure}

Let us analyze the transport through various modes as a function of the wave
frequency. As an example, the measured frequency dependence of the mode-transport
coefficient [Eq.~\eqref{eq:Tnm-def}] for the third $T_{3}(\nu)$ and the fourth
$T_{4}(\nu)$ modes is depicted in Fig.~\ref{fig:emptywaveguide_mode} by solid
curves. There are several interesting features in this figure. First of all, there
is a pronounced spike in $T_{3}(\nu)$ at the cut-off frequency
$\nu_c^{(3)}=4.5$\,GHz, at which the third mode opens and starts to propagate. The
experimental value of $\nu_c^{(3)}$ coincides very well with the theoretical
prediction, $\nu_c^{(3)}=3c/2w$ [see Eq.~\eqref{eq:cutoff} and comments below it].
Above this critical value the coefficient $T_{3}$ decreases for increasing
frequency.

The frequency dependence of $T_{4}(\nu)$ looks quite similar to that of
$T_{3}(\nu)$, apart from the shift of the spectrum by the value
$\nu_c=c/2w=1.5$\,GHz. This is because the propagation of the fourth mode emerges
above the cut-off frequency $\nu_c^{(4)}=4c/2w=6.0$\,GHz. In general, all
mode-transport coefficients $T_n(\nu)$ have somewhat similar frequency dependence.
Namely, they manifest the corresponding spikes at the corresponding cut-off
values, $\nu_c^{(n)}=nc/2w$, and decrease with a further frequency increase.

In order to understand such specific frequency dependence of the coefficients
$T_n(\nu)$, one has to use Eqs.~\eqref{eq:Snm-dg} and \eqref{kn}. Since
in an empty waveguide all propagating modes are independent of each other, the
mode-transport coefficient reads
\begin{eqnarray}\label{eq:Tn-EmW}
  T_n(\nu)&=&|S_{nn}|^{2}= \frac{|A_n|^2}{4|k_n|^2}\Theta(N_w-n)\nonumber\\[6pt]
          &=&\frac{c^2|A_n|^2}{16\pi^2(\nu^2-n^2\nu^2_c)}\Theta(\nu-n\nu_c).
\end{eqnarray}
One can easily achieve a nice correspondence of this theoretical result to the
experimental data by a simple fit of the factor $A_n$. To do this, we must also
take into account an evident assumption that the wave number $k$,
(i.e., the frequency $\nu$) has a small imaginary part $\textrm{Im}\,k$ (or
$\textrm{Im}\,\nu$), which emerges because of an absorption in the waveguide.

Typically, the absorption is frequency dependent, but in the analysis it is
sufficient to treat it as a constant for each mode, i.\,e., to regard
$\textrm{Im}\,k$ as a function of the mode index $n$ only. In
Fig.~\ref{fig:emptywaveguide_mode} the theoretical expression \eqref{eq:Tn-EmW} is
displayed by dotted curves. Note, however, that the fast oscillations that are
clearly seen in the experimental curves, are not described by our theory. These
oscillations are caused by multiple wave reflections from the waveguide ends
(occurring because of non-perfect absorbers) and from the antennas.

The frequency dependence experimentally observed for the coefficient $T_n(\nu)$ is
explained by the presence of the squared modulus of longitudinal wave number $k_n$
in the denominator of Eq.~\eqref{eq:Tn-EmW}. Apart from the frequency dependence,
this term, $|k_n|^2=\frac{4\pi^2}{c^2}(\nu^2 - n^2\nu_c^2)$, provides the global
dependence of $T_n(\nu)$ on the mode index $n$.
One can see that at fixed frequency $\nu$ the higher the mode index
$n$, the larger the coefficient $T_n(\nu)$. That is why the highest propagating
mode with $n=N_w$ always gives the main contribution to the total transport
coefficient $T$, especially, in the vicinity of its cut-off,
$\nu_c^{(N_w)}=N_wc/2w$. This effect is displayed in
Fig.~\ref{fig:emptywaveguide_snm}.

\section{Periodic Structure}

The features of the empty waveguide discussed above, were explored as a necessary
precondition for our further analysis. As a second step, we studied a periodic
structure without any disorder. To this end, the brass bars of thickness
$d_b=8$\,mm measured along the $x$ axis, were inserted into the waveguide as is
shown in Fig.~\ref{fig:sketch}. The constant spacing $d_a$ between neighboring
bars was chosen as $d_a=17$\,mm and $d_a=32$\,mm for two different realizations,
yielding a period $d=d_a+d_b=25$\,mm for the first realization with 42 bars, and
$d=40$\,mm for the second one with 26 bars. The measurements were performed as
before, giving rise to the scattering matrix $S_{nm}$ of the periodic setup.
Both periodic structures gave similar results; therefore, in what follows, we
restrict the discussion to the second case only. It would be interesting to investigate the
behavior of the bands and the correlation gaps with decreasing number of
scatterers $N$, but this is beyond the scope of the present paper.

\begin{figure}
\includegraphics[width=\columnwidth]{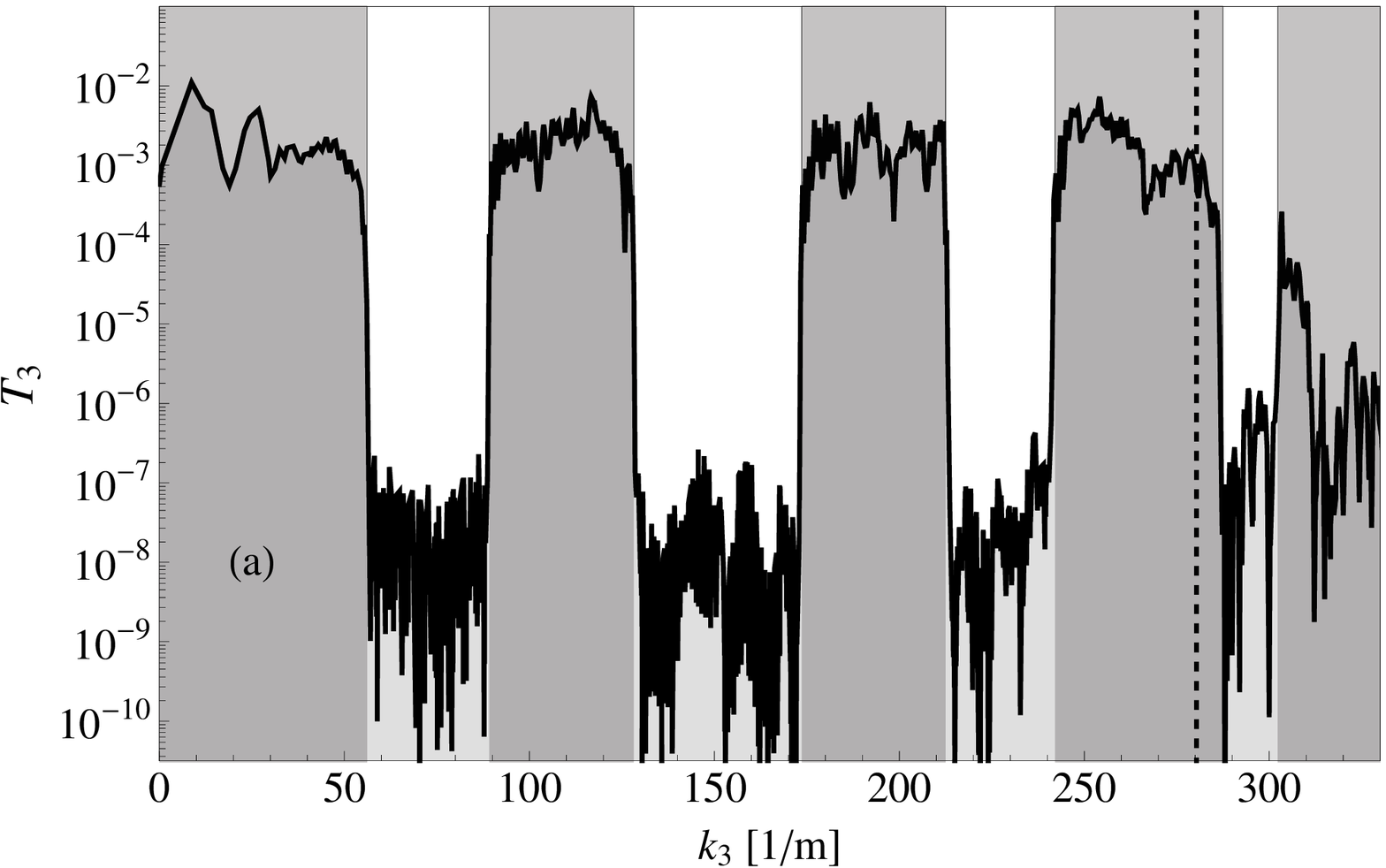}
\includegraphics[width=\columnwidth]{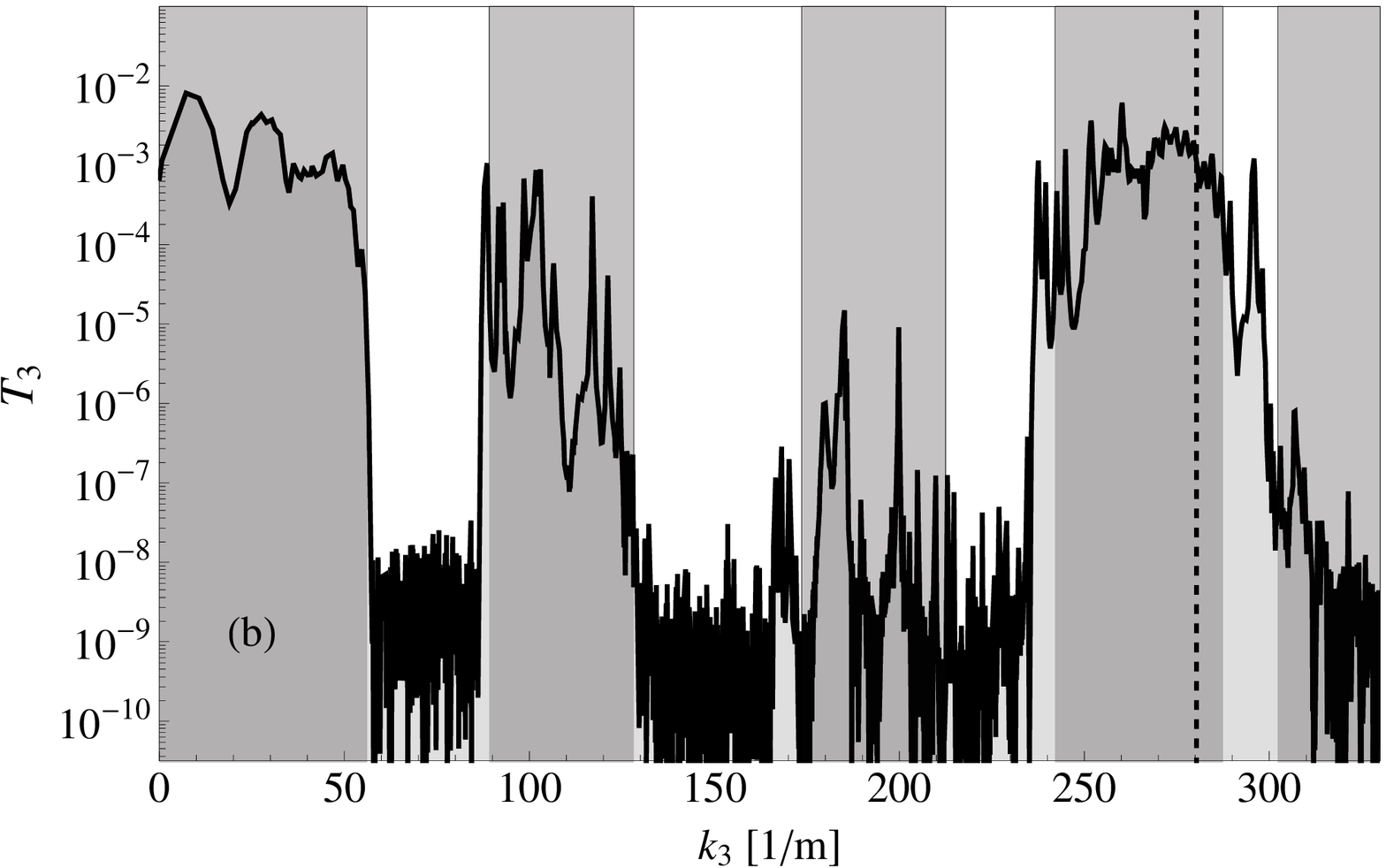}
\includegraphics[width=\columnwidth]{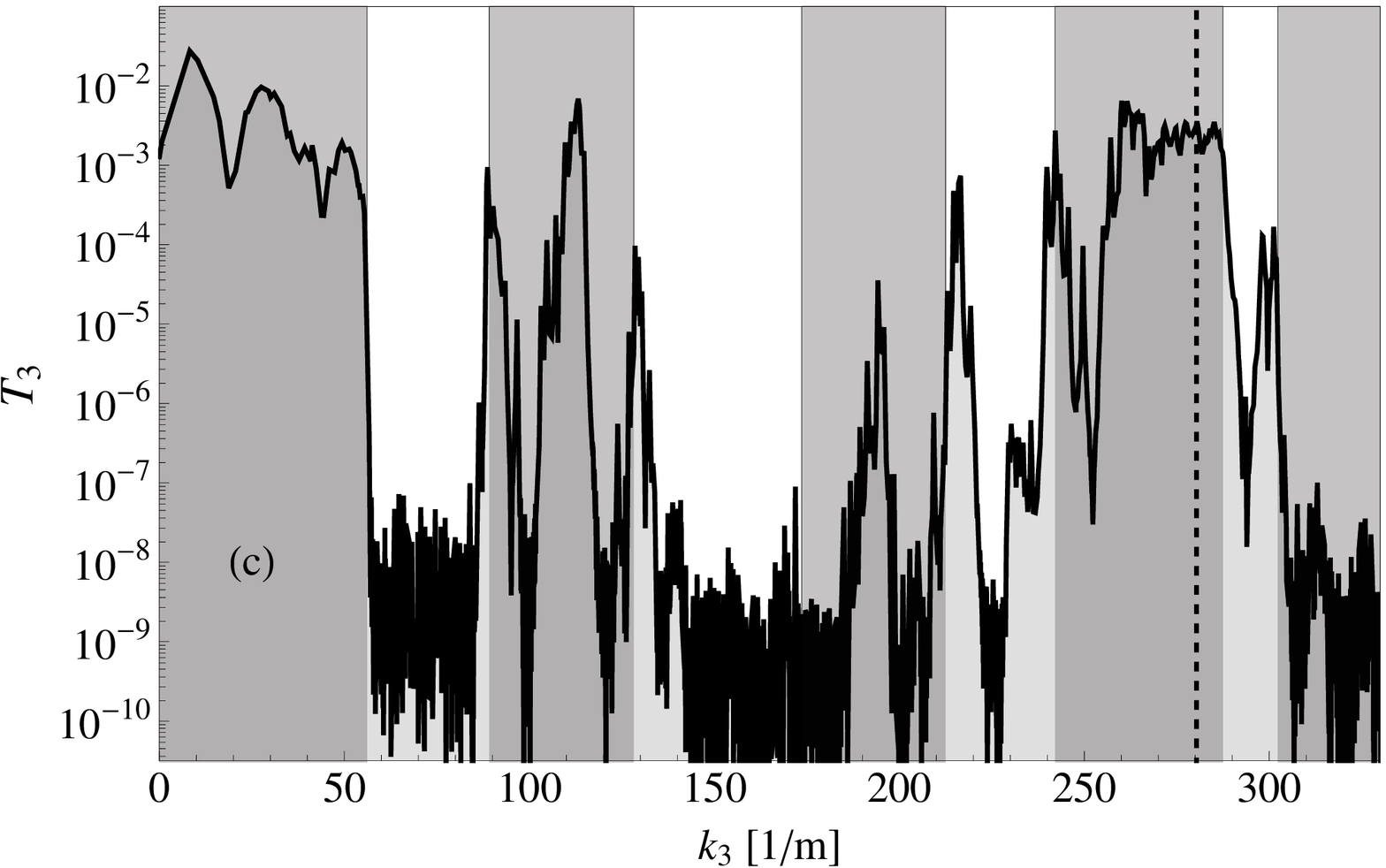}
\caption{\label{fig:periodic} (Color online)
Band structure for (a) periodic (b) white noise and (c) correlated disorder
arrangement with $\langle d_a(j)\rangle=d_a=32$\,mm.  Shaded regions indicate the
bands restricted by phenomenologically obtained band edges (vertical lines).
Dashed vertical line indicates the Fabry-P\'{e}rot resonance (see Section
\ref{sec:white-noise}).}
\end{figure}

Since the periodic structure is arranged along the wave propagation (along the
$x$ axis) and the bars are homogeneous along the $z$ axis, the propagating modes
remain independent as in the empty waveguide. Indeed, the scattering matrix
[Eq.~\eqref{eq:Snm-def}] was experimentally found to be diagonal within the same
accuracy as before. As a consequence, the spectrum of the mode-transport
coefficient $T_n(\nu)=|S_{nn}|^2$ looks similar for all propagating modes,
provided the shift of spectrum by particular cut-off frequencies,
$\nu_c^{(n)}=nc/2w$ is done. This means that $T_n(\nu)$ is convenient to represent
as a function of the corresponding longitudinal wave number $k_x=k_n$
[see Eq.~\eqref{kn}]. Fig.~\ref{fig:periodic}(a) shows the semilogarithmic
experimental plot of the mode-transport coefficient $T_3(k_3)$ for the third
propagating mode. One can clearly see a band structure in the dependence
$T_3(k_3)$, as is expected for a periodic arrangement. Note that the experimental
transmission in the gaps is smaller than in the bands over several orders of
magnitude.

From the physical point of view our setup with periodic filling of brass bars
represents a set of $N_w$ independent 1D periodic arrays of two alternating $a$
and $b$ layers. Note, however, that each unit $(a,b)$ cell of every $n$th array
has a quite complicated structure. While the first $a$-layer is a trivially
homogeneous air spacing of thickness $d_a$ with the wave number $k_n$, the second
$b$-layer of thickness $d_b$ is not homogeneous along the $y$-axis. Indeed, since
the height ($5$\,mm) of the brass bar is smaller than the waveguide height
($h=8$\,mm), there is an air slot between the metallic bar and the waveguide top
plate, see Fig.~\ref{fig:sketch}. For this reason, in our analysis we shall
characterize the $b$-layers by the wave number $k_b=n_bk_n$, thus, introducing an
{\it effective} refractive index $n_b$ that should be fixed by the experimental
data. In this way, in order to describe the transmission band structure for every
$n$th array of bilayers (for every $n$th propagating mode) we employ the effective
Kronig-Penney model with the common dispersion relation for the Bloch wave number
$\kappa$ (see, e.g., Refs.~\onlinecite{kro31,mar08}),
\begin{eqnarray}\label{eq:dispersion}
\cos(\kappa d)&=&\cos(k_nd_a)\cos(n_bk_nd_b)\nonumber\\
               &&-\alpha_+\sin(k_nd_a)\sin(n_bk_nd_b).
\end{eqnarray}
For the standard bilayer
models the factor $\alpha_+$ is the arithmetic average of the impedance ratio and
its inverse value, $\alpha_+=(Z_a/Z_b+Z_b/Z_a)/2$. Since the impedance $Z_b$ of
$b$-layers is unknown for our system, the factor $\alpha_+$ is regarded as the
second fitting parameter. Note that $\alpha_+$ equals one for the case of two basic layers
$a$ and $b$ which are optically matched ($Z_a=Z_b$).

The solution of Eq.~\eqref{eq:dispersion} gives the Bloch wave number
$\kappa=\kappa(k_n)$ as a function of the mode wave number $k_n$ (or, as a
function of the wave frequency $\nu$ and the mode index $n$). The real values of
$\kappa$ are achieved for such $k_n$ for which $|\cos(\kappa d)|<1$, thus,
resulting in the emergence of spectral {\it transmission bands}. Otherwise, for
$k_n$ for which $|\cos(\kappa d)|>1$, the Bloch wave number $\kappa$ turns out to
be imaginary leading to the emergence of {\it spectral gaps}. The dispersion
relation \eqref{eq:dispersion} gives the band edges at the points where
$\cos(\kappa d)=\pm1$.

\begin{figure}
\includegraphics[width=\columnwidth]{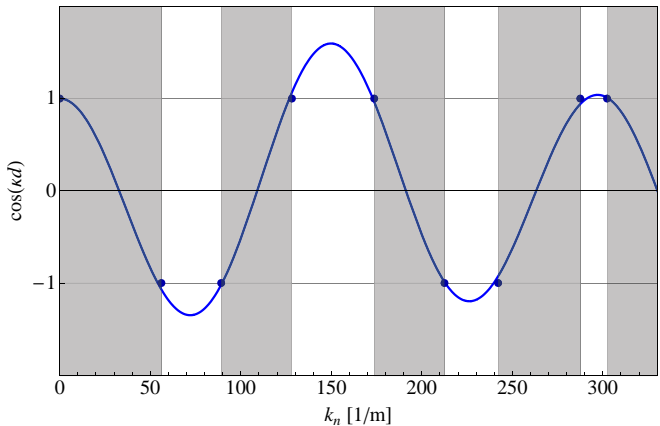}
\caption{\label{fig:bandstr} (Color online)
The r.h.s.\,of the dispersion relation \eqref{eq:dispersion}  (solid line) for the best
fit of the parameters. The spacing is $d_a=32$\,mm with fitting parameters
$\alpha_+=1.6$  and $n_b=1.42$. Full circles correspond to the band edges
extracted from Fig.~\ref{fig:periodic}(a).}
\end{figure}

As a first step, we manually extracted the positions of the band edges from the
measured spectrum of the coefficient $T_n(\nu)=|S_{nn}|^2$, see vertical lines in
Fig.~\ref{fig:periodic} for these positions. After, we fitted the dispersion
relation \eqref{eq:dispersion} by changing the values of $\alpha_+$ and $n_b$, and
keeping the crossing points of its r.h.s. with the horizontal lines $\cos(\kappa
d)=\pm1$, in the correspondence with the found band edges. The result is displayed
in Fig.~\ref{fig:bandstr} for $\alpha_+=1.6$ and $n_b=1.42$. The dark dots in the
figure show the extracted band edges.

\section{Positional Disorder}

In order to study the effect of disorder, we shift each brass bar by a random
value along the waveguide length (the $x$-axis), keeping the average
distance between the bars constant. Since the disorder is a function of the
$x$-coordinate only, the propagating modes remain independent as in the considered
above arrangements \cite{izr04a,izr05a}. In such a manner, we can consider every $n$th
channel as an independent one-dimensional array of two alternating $a$ (air
spacing) and $b$ (brass bar) layers. Theoretically, the positional disorder is
incorporated via the random thickness of $a$-layers only,
\begin{equation}\label{eq:da-j}
d_a(j)=d_a+\varrho_a(j)\,,\qquad \langle d_a(j)\rangle=d_a\,,
\end{equation}
while the thickness $d_b$ of $b$ layers is constant.
The integer $j=1,2,3,\ldots,N_d$, with $N_d=\llbracket L/d\rrbracket$
enumerates the unit $(a,b)$ cells, $d_a$ is the average thickness of the $a$ layer;
and $\varrho_a(j)$ stands for small random variations. Note that in the absence of
disorder ($\varrho_a=0$) the set of 1D bilayer arrays represents the periodic
structure with the period $d=d_a+d_b$ that was considered in previous section.

The random sequence $\varrho_{a}(j)$ is statistically homogeneous with the zero
average and given variance $\sigma_a^2$,
\begin{equation}\label{eq:CorrDef1}
\langle\varrho_{a}(j)\rangle=0\,,\qquad \langle\varrho_a^2(j)\rangle=\sigma_a^2\,.
\end{equation}
The binary correlator is defined by
\begin{equation}\label{eq:CorrDef2}
\langle\varrho_a(j)\varrho_a(j')\rangle= \sigma_a^2K_a(j-j')\,.
\end{equation}
Averaging $\langle\ldots\rangle$ is
performed over the whole array of layers or over different ensembles, which are
assumed to be equivalent. As one can see, the two-point correlator $K_{a}(j-j')$
is normalized to unity, $K_{a}(0)=1$.

If the positional disorder is weak ($k^2_a\sigma_a^2\ll 1$), all transport
properties are determined entirely by the randomness power spectrum
$\mathcal{K}_a(k)$,
\begin{subequations}\label{eq:FT-K}
\begin{eqnarray}
\mathcal{K}_a(k)&=& 1+2\sum_{r=1}^{\infty}K_a(r)\cos(kr),\\[6pt]
K_a(r)          &=& \frac{1}{\pi}\int_{0}^{\pi}dk\mathcal{K}_a(k)\cos(kr).
\end{eqnarray}
\end{subequations}

Since the correlator $K_{a}(r)$ is a real and even function of
the difference $r=j-j'$ between cell indices, its Fourier transform
$\mathcal{K}_a(k)$ is a real and even function of the dimensionless wave number $k$.
According to the Wiener-Khinchin theorem, the power spectrum $\mathcal{K}_a(k)$
is non-negative for {\it any} real sequence $\varrho_{a}(j)$.

The transport through any 1D disordered structure is governed by the phenomenon of
Anderson localization (see, e.g., Ref.~\onlinecite{lif88}). Its principal
concept is that all transport characteristics depend only on the ratio between the
system length $L$ and localization length $L_{loc}$. Thus, the localization length
is the key quantity that controls the transport in a 1D geometry. The analytical
expression for the localization length in the periodic-on-average model with
bilayer structure was derived in Ref.~\onlinecite{mak07c}. Specifically, the case
of weak disorder in the thickness of only one kind of layers was analyzed, which
is close to our experimental setup.

However, our model is essentially quasi-one-dimensional. On the other hand, in
Refs.~\onlinecite{izr04a,izr05a} it was shown that in quasi-1D waveguides with random
stratification {\it along} the propagation, each of the (independent) modes can be
associated with their own length $L_{loc}(k_n)$. The latter can be obtained by a
direct substitution of specific mode characteristics into the general expression
for the localization length of 1D random structures.
Applying the results from Refs.~\onlinecite{izr04a,izr05a,mak07c}, one can
readily get the following expression for the inverse localization length,
associated with the $n$th propagating mode of our setup,
\begin{equation}\label{eq:Lloc}
L_{loc}^{-1}(k_n)= \frac{\alpha_-^2k_n^2\sigma_a^2}{2d\sin^2(\kappa d)} \mathcal{K}_a(2\kappa
d)\sin^2(n_bk_nd_b).
\end{equation}
Here $\alpha_-^2$ is the so-called mismatching factor.
For standard bilayer models $\alpha_-$ is related to
$\alpha_+$ in the dispersion relation [Eq.~\eqref{eq:dispersion}] by the equality
$\alpha_+^2-\alpha_-^2=1$. Therefore, for the perfect matching for which
$\alpha_+=1$, we have $\alpha_-=0$ resulting in a divergence of the localization
length, $L_{loc}=\infty$. Since for our system $\alpha_+=1.6$, the mismatching
factor is $\alpha_-^2=1.56$. Note that Eq.~\eqref{eq:Lloc} is a particular case of
a more general expression derived in Ref.~\onlinecite{izr09} for the case of both
$a$ and $b$ positional disorders (when thicknesses of both $a$ and $b$ layers are
randomly perturbed).

\begin{figure}
\includegraphics[width=\columnwidth]{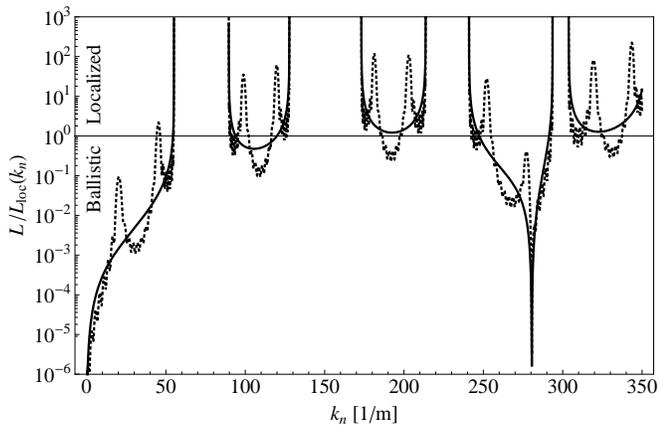}
\caption{\label{fig:oneoverln}
Ratio $L/L_{loc}(k_n)$ formed by Eq.~\eqref{eq:Lloc} for white-noise disorder
(solid curve) and for correlated disorder (dotted curve). Parameters are
$d_a=32$\,mm, $\sigma_a=2.21$\,mm, $\alpha_-^2=1.56$. Horizontal line
indicates the equality $L=L_{loc}(k_n)$.}
\end{figure}

\subsection{White-noise disorder}
\label{sec:white-noise}

In Fig.~\ref{fig:oneoverln} the inverse localization length $L_{loc}^{-1}(k_n)$ of
the $n$th propagating mode, measured in the units of the waveguide length
$L=1.5$\,m, is plotted against $k_n$. The 26 displacements $\varrho_a(j)$
were drawn randomly from the interval -5\,mm to 5\,mm, while the variance was constrained to
$\sigma_a^2=4.9\,\mathrm{mm}^2$. The ratio of $\sigma_a$ and the spacing $d_a$ is
about 7\%.

Now we start with the discussion of experimental data for the case of uncorrelated
(white-noise) disorder. In this case the binary correlator is
$K_{a}(j-j')=\delta_{jj'}$, and consequently, the power spectrum is constant,
$\mathcal{K}_a(k)=1$. The corresponding ratio $L/L_{loc}(k_n)$ is presented in
Fig.~\ref{fig:oneoverln} by solid curve. One can see that within the first
spectral band the inverse localization length monotonously increases from zero to
infinity dependening on $k_n$. Within any other Bloch band the mode-localization
length $L_{loc}(k_n)$ vanishes at the band edges (where $\kappa d=0,\pi$) because of
the term $\sin^2(\kappa d)$ in the denominator of Eq.~\eqref{eq:Lloc}.

It is important that the mode-localization length $L_{loc}(k_n)$ exhibits the
Fabry-P\'{e}rot resonances associated with multiple wave-reflections inside every
$b$ layer from its interfaces. As is known, the resonances appear when the constant width
$d_b$ is equal to an integer multiple of half of the wavelength inside the layer:
\begin{equation}\label{eq:FabryPerot}
k_n=\pi s/n_bd_b\,,\qquad s= 1,2,3,\ldots.
\end{equation}
At the resonances the factor $\sin^2(n_bk_nd_b)$ in
Eq.~\eqref{eq:Lloc} vanishes, resulting in resonance divergence of the
localization length and, consequently, in suppression of the localization (the
$n$th mode becomes fully transparent). Remarkably, the Fabry-P\'{e}rot resonance
should be quite broad, since it is caused by the vanishing of a smooth trigonometric
function. This effect was analyzed both theoretically and experimentally in
Refs.~\onlinecite{lun09a,lun09b}. Since in our experimental setup $n_b=1.42$ and
$d_b=8$\,mm, the first ($s=1$) Fabry-P\'{e}rot resonance is expected at
$k_n=280.5\textrm{\,m}^{-1}$ and this is what is observed in the fourth transmission band in
Fig.~\ref{fig:oneoverln}.

In accordance with single-parameter scaling, there are only two transport
regimes in the 1D Anderson localization: the ballistic ($L\ll L_{loc}$) and
localized ($L_{loc}\ll L$) regimes. In the ballistic regime a sample is
practically transparent, and in the localized transport a 1D disordered structure
almost perfectly reflects classical or quantum waves.

In Fig.~\ref{fig:oneoverln} the horizontal straight line  $L/L_{loc}(k_n)=1$
separates the upper region, $L/L_{loc}(k_n)>1$ (where the $n$th propagating mode
is localized and therefore closed), from the lower one, $L/L_{loc}(k_n)<1$ (where
the $n$th propagating mode is transparent). As one can see from
Fig.~\ref{fig:oneoverln}, in our experiments the first and fourth spectral bands
should be almost completely transparent. Note that the transparency of the first
band is due to a very weak influence of the disorder, while the transparency of
fourth band is caused by the broad Fabry-P\'{e}rot resonance.

The predicted features of the system are clearly displayed in
Figs.~\ref{fig:periodic} for the experimentally measured mode-transport
coefficient $T_3(k_3)$. The comparison of Figs.~\ref{fig:periodic} (a) and (b)
shows that the introduced disorder does not reduce transmission in the first band,
where $k_n<50\textrm{\,m}^{-1}$. In the fourth band between $240\textrm{\,m}^{-1}$
and $290\textrm{\,m}^{-1}$ again there is only a weak influence of disorder.
Only in the second, third and fifth bands the compositional disorder
reduces the mode transport by 2--3 orders of magnitude.

\begin{figure}
\includegraphics[width=\columnwidth]{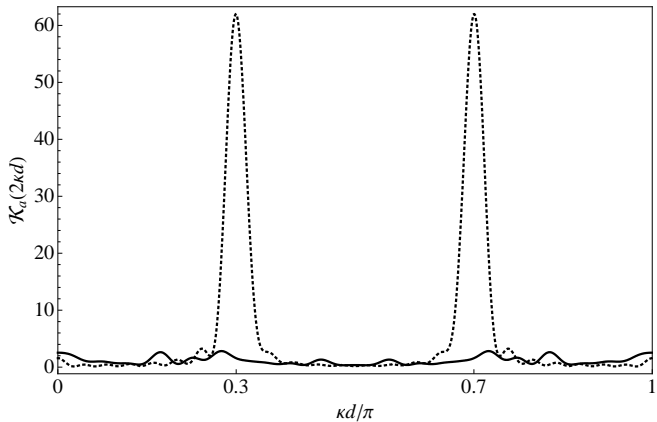}
\caption{\label{fig:ftofc}
Power spectrum $\mathcal{K}_a(2\kappa d)$ versus dimensionless Bloch phase $\kappa d/\pi$ for
correlated (dotted curve) and uncorrelated (solid curve) random sequences. The
factor $\mathcal{K}_a(2\kappa d)$ is calculated from Eq.~\eqref{eq:FT-K} taking
only the first 13 terms in the sum that corresponds to an arrangement with
$N_d=\llbracket L/d\rrbracket=26$ unit $(a,b)$ cells with $\langle
d_a(j)\rangle=d_a=32$\,mm.}
\end{figure}

\subsection{Correlated disorder}

As one can see from Eq.~\eqref{eq:Lloc}, the inverse localization length
$L_{loc}^{-1}(k_n)$ follows the profile of the power spectrum
$\mathcal{K}_a(2\kappa d)$. This fact is of special interest in view of
experimental realizations of positional disorder with specific long-range
correlations. In particular, one can artificially construct an array of random
bilayers with such a power spectrum that vanishes abruptly within prescribed
intervals of the mode wave number $k_n$, resulting in the divergence of the
mode-localization length. Thus, by a proper design of the positional disorder, one
can create quasi-1D bilayer structure with a {\it selective} mode transmission
$T_n(k_n)$ in dependence on the frequency $\nu$.

To do this, we perturbed all periodically spaced $N_d=26$ brass bars by random
values $\varrho_a(j)$ ($j=1,2,3,\ldots,N_d$) correlated in such a way that the
power spectrum $\mathcal{K}_a(2\kappa d)$ is zero everywhere, except for narrow
regions around $\kappa d/\pi=0.3$ and $0.7$.

We can solve the problem of generating the random sequences with prescribed
short- or long-range correlators by employing a widely used convolution method
originally proposed by Rice~\cite{ric54b}.
The details of various applications of this method can be found in
Refs.~\onlinecite{izr99,kro99,izr01a,izr01b,izr03d,izr05a,kuh00a,kuh08a,izr04a,wes95,czi95,mak96,rom99,gar99,cak06,izr07,apo08}.
With the use of this method, the resulting curve for $\mathcal{K}_a(2\kappa d)$
has the form shown in Fig.~\ref{fig:ftofc}. For comparison, the power spectrum for an
uncorrelated random sequence is also shown. The inverse dimensionless localization
length $L/L_{loc}(k_n)$ that corresponds to the specified correlated disorder, is
depicted in Fig.~\ref{fig:oneoverln} by the dotted curve. In accordance with this
curve, one should expect the appearance of correlation gaps, i.e.,
additional gaps due to correlated disorder (namely, one correlation gap in the
first band and two in the second and third transmission bands). In the first band
the first possible correlation gap at $k_n\approx25\textrm{\,m}^{-1}$ remains in the ballistic
regime and can not be observed.  The same happens in the fourth band, where only
the first correlation gap can be observed. Because of the overlap with the Fabry-P\'{e}rot
resonance the second correlation gap is shifted to the ballistic regime.

Fig.~\ref{fig:periodic}(c) shows the mode-transmission through the correlated
random arrangement. Globally, the spectral structure looks similar to that for the
uncorrelated random setup. Indeed, we can see that the transport decreases with
increase of $k_n$, with an exception of the fourth band containing the Fabry-P\'{e}rot
resonance.
In the first band the mode-transport coefficient $T_n(k_n)$ is comparable to that
for both regular and uncorrelated random setup: Figs.~\ref{fig:periodic}(a)
and (b). This is because in this band the localization length is larger than the
system size, $L\ll L_{loc}(k_n)$. Only in the region of the correlation gap at
$k_n=40\textrm{\,m}^{-1}$ does the localization length shrink, such that $L_{loc}(k_n)\approx L$,
resulting in an very weak, but recognizable correlation gap. This correlation gap
is an interesting special case, because it is in the regime intermediate
between ballistic and localized.

The second band exhibits two {\it correlation gaps} that are clearly seen, in
spite of the fact that in this band the transport is reduced by two orders of
magnitude. In the third spectral band, at about $k_n=200\textrm{\,m}^{-1}$, the transport
already decreased to noise level. However, in the fourth band containing the
Fabry-P\'{e}rot resonance, the transport is quite large and comparable to that for the
regular arrangement -- apart from one correlation gap. Finally, within the fifth
band the mode-transport coefficient is reduced by randomness so strongly that it
is not possible to reveal any difference between correlated and uncorrelated
disorder. All these properties are in a complete agreement with the predictions
provided by our theoretical model.

\begin{figure}
\centering
\includegraphics[width=\columnwidth]{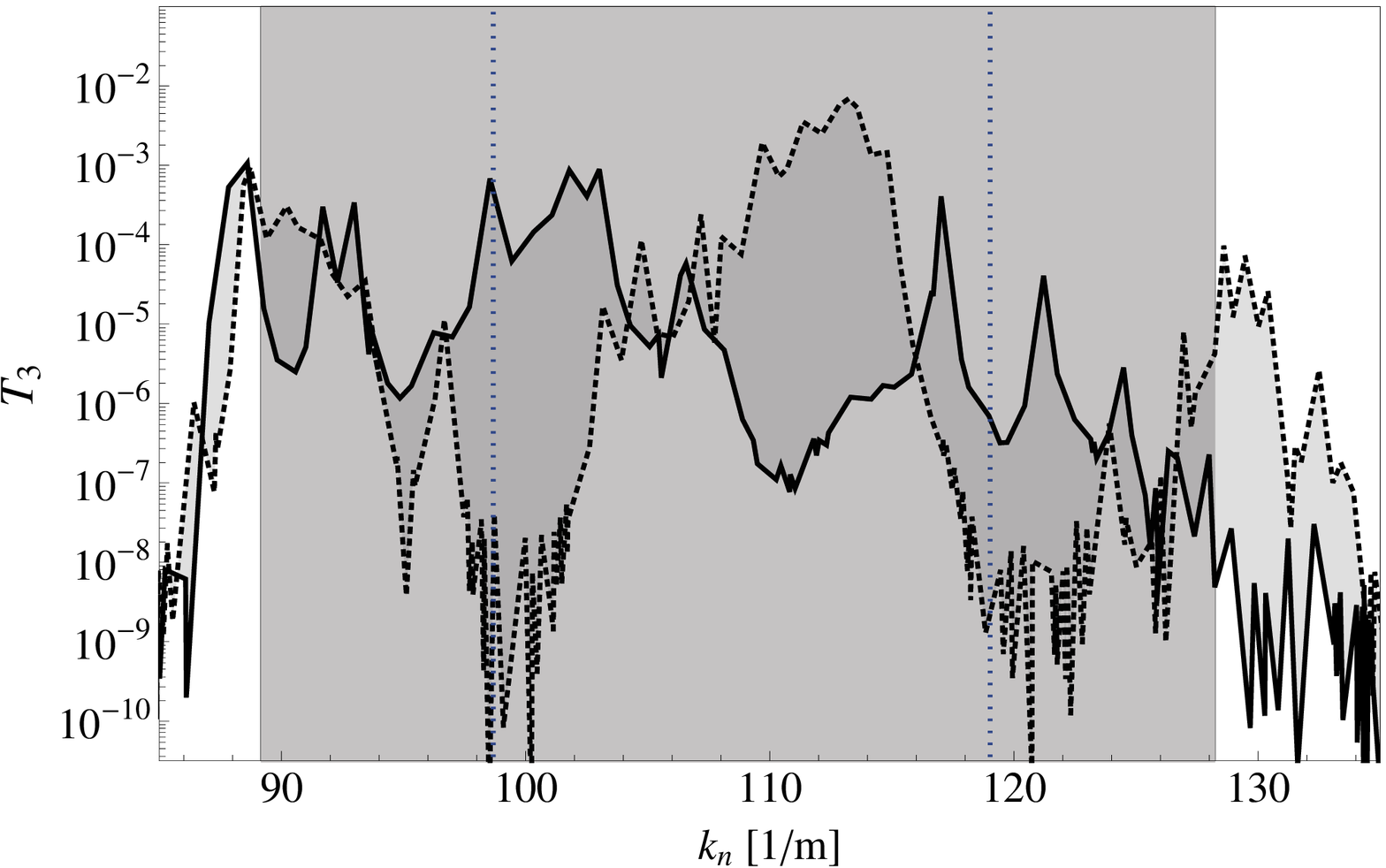}
\includegraphics[width=\columnwidth]{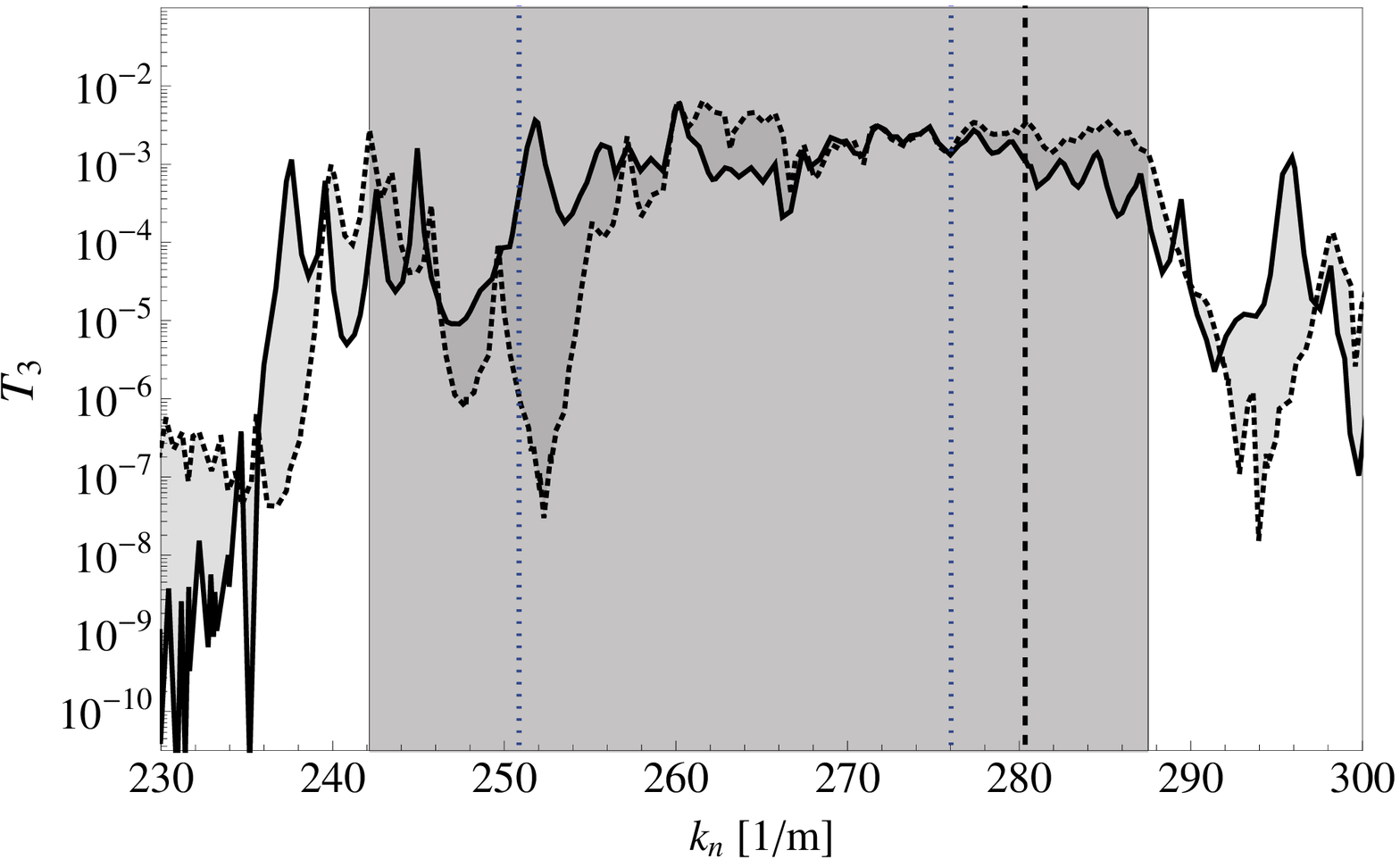}
\caption{\label{fig:bandstrcorr} (Color online)
Comparison of the mode-transport through uncorrelated (solid curve) and correlated
(dotted curve) random  arrangements. Upper figure shows the second band;
lower figure presents the fourth (Fabry-P\'{e}rot) band. Vertical dotted line
indicates expected correlation gaps for $\kappa d/\pi=0.3, 0.7$. Vertical dashed
line indicates the position of the Fabry-P\'{e}rot resonance at $k_n=280.5\textrm{\,m}^{-1}$.}
\end{figure}

In order to compare the effects of uncorrelated and correlated disorder in more
detail, we present Fig.~\ref{fig:bandstrcorr}, which illustrates the second and fourth
spectral bands of the mode-transport coefficient $T_n(k_n)$ for the third
propagating mode. The expected positions of the correlation gaps at $\kappa
d/\pi=0.3, 0.7$ are indicated by vertical dotted lines, and the position of the
Fabry-P\'{e}rot resonance is given by the vertical dashed line.

In the second spectral band (see upper part of Fig.~\ref{fig:bandstrcorr}), the
correlation gaps are easily recognized, showing the variation of the amplitude
over five orders of magnitude. There is good agreement between the theoretically
predicted and experimentally found positions of these correlation gaps. Between
the correlation gaps, i.e., in the vicinity of $k_x=110\textrm{\,m}^{-1}$, the mode-transport
coefficient for the correlated disorder is two orders of magnitude higher than for
uncorrelated disorder. Again, this is not surprising, since the power spectrum
$\mathcal{K}_a(k)\approx 1$ for uncorrelated disorder, whereas in the case of
correlated disorder it is close to zero outside the correlation gaps.
Remember that the mode-localization length [Eq.~\eqref{eq:Lloc}] tends to infinity as
$\mathcal{K}_a(k)\to0$.

In the fourth spectral band containing the Fabry-P\'{e}rot resonance (lower part of
Fig.~\ref{fig:bandstrcorr}) only the first correlation gap is found. The second
correlation gap is at a short distance from the Fabry-P\'{e}rot resonance and
therefore, is suppressed. Indeed, here because of the resonance the mode-localization
length is still quite large, $L\ll L_{loc}(k_n)$, in order to contribute to
transmission of the waveguide of finite length, see also Fig.~\ref{fig:oneoverln}.
Only for the first correlation gap does the mode-localization length turn out to be
sufficiently small, $L_{loc}(k_n)\ll L$, to observe this gap. Note that for
white-noise disorder the fourth band is almost fully transparent exclusively because
of the Fabry-P\'{e}rot resonance. Nonetheless, the specific correlations in the
positional disorder of the brass bars can induce a correlation gap provided that the
distance between the resonance and the corresponding maximum of the power spectrum
$\mathcal{K}_a(2\kappa d)$, are sufficiently long.

Finally, we would like to stress that all peculiar effects discussed above, are
experimentally observed for a quite short quasi-1D system, with as few as $N_d=26$
brass bars. This fact confirms the expectation that the analytical
expression for the localization length, formally obtained for infinite systems, is
actually applicable for real experimental situations with a relatively small
number of scatterers.

\section{Conclusions}
We studied transport through a quasi-1D microwave waveguide
with periodic and randomized scattering potentials.
The potentials were stratified, i.e., homogeneous along the width of the
waveguide. In this case the propagation of waves occurs independently
along each open channel, thus allowing one to apply the theory of
1D transport to any partial channel and make conclusions about both partial and total transmission.

In the experimental setup the incoming and outgoing waves are induced by pointlike
antennas. Therefore, one of the problems was to understand the influence of the
coupling of antennas to the waveguides, as well as the role of reflection from the
antennas and the open ends. Also, the effect of absorption in the walls influenced
the transport characteristics. We have tried to incorporate all these effects in
our phenomenological theory aiming to explain the observed data.

Main attention was paid to selective transport due to long-range correlations in
the disorder.We generated the disorder by intentionally shifting the position of
bulk scatterers.
In order to quantify our experimental setup, we started with the case of an open
waveguide without any scatterers. In this case we have confirmed that the coupling
between various channels due to antennas is small and can be neglected when
compared to the theoretical predictions. Then we studied scattering
through the waveguide with periodically inserted scatterers chosen in the form of
finite-thickness brass bars. With such an arrangement, the transmission spectrum was
found to have a band structure, as expected. However, in order to analytically
describe the structure of the spectrum, one must create a phenomenological
model that properly corresponds to the experimental setup.

The specific problem we encountered was that the periodic brass bars
inserted in the waveguide cannot simply be considered as finite-thickness barriers with
certain refractive indices. They were made of metal, and in order to allow
propagation they have a finite height that is less than the vertical dimension of the
waveguide. In this case it impractically to rigorously create a
relatively simple and adequate theoretical model, and instead we decided to use
the standard dispersion relation known in the theory of 1D models, by
introducing two fitting parameters. In this way we were able to give an analytical
expression for the positions of transmission bands and spectral gaps. This
expression helped us describe another experiment, in which the
positions of brass bars were slightly perturbed, both in a random manner and with
specific long-range correlations.

For the case of uncorrelated disorder, we obtained experimental data that can be
described by an analytical expression for the localization length.
For any kind of correlation, the corresponding expression was derived in
Refs.~\onlinecite{izr05a,izr04a,mak07c,izr09} and one of our interests was to
check how well our experimental data correspond to the predictions
of the theory. It should be stressed that this problem is far from trivial,
since the expression for the localization length was developed for the infinite
number of scatterers and for a weak disorder. In our case, the number of scattering
bars $N_d=26$ was relatively small, and many other experimental imperfections
(absorbtion, non perfect coupling, etc.) are not taken into account in the
expression for the localization length. However, the experimental data were found
to correspond well with our expectations following from the theory. In
particular, a quite unusual effect of Fabry-P\'{e}rot resonances was observed in
experimental data, which is fully explained by the theory.

The Fabry-P\'{e}rot resonances were predicted to play a specific role in the transport
(see discussion in Refs.~\onlinecite{lun09a,lun09b}). Namely, if one such
resonance emerges inside the transmission band, the transmission coefficient
turns out to be very large as compared with the bands without such a resonance.
The nontrivial point is that the resonance effect occurs not only for a specific
frequency, but in a large interval of frequencies inside the band. In our
experiment we have clearly observed the Fabry-P\'{e}rot resonance in the fourth band.
The transport in this band would be negligible otherwise.

Another specific interest was to check whether in our setup with a finite number
of brass-bar scatterers one can observe the effect of long-range correlations
embedded intentionally in the positions of bars. According to the theory, with
specific long-range correlations one can construct such potentials for which the
localization length (obtained in the first order perturbation theory) diverges (or
very small) in finite windows of frequency. In these windows one can observe
either strongly enhanced or strongly suppressed transport, compared to the white-noise
disorder case. In order to observe experimentally the effect of suppression of the
transmission, we created disordered perturbations of the positions of brass bars,
in accordance with theoretical expressions for long-range correlations.
Remarkably, our experimental data clearly manifest the emergence of new gaps
inside the transmission bands, in good correspondence with analytical expressions.
In particular, these gaps were found to arise at the positions that were predicted
theoretically.

\begin{acknowledgments}
The authors are thankful to Timur Tudorovskiy for fruitful discussions.
O.D., U.K., H.-J.S., and F.M.I acknowledge support from the
Deutsche Forschungsgemeinschaft within the research group 760
``Scattering Systems with Complex Dynamics''.
F.M.I acknowledges additional support from
Vicerrector\'{\i}a de Investigaci\'{o}n y Estudios de Posgrado Grant
No. EXC08-G of the Benem\'{e}rita Universidad Aut\'{o}noma de Puebla (Mexico).
\end{acknowledgments}

\bibliography{paperdef,paper,book,newpaper}

\end{document}